\documentclass[aps,pra,reprint,superscriptaddress,amsmath,amssymb,floatfix]{revtex4-1}
\usepackage{epsfig}
\usepackage[colorlinks=true,citecolor=blue,linkcolor=blue,urlcolor=blue]{hyperref}
\usepackage{siunitx}
\usepackage{nicefrac}
\usepackage{tabularx}
\usepackage{apjfonts}
\usepackage{booktabs}
\usepackage{multirow} 

\newcommand{\apjl}{Astrophys.\ J. Lett.}
\newcommand{\aap}{Astron.\ Astrophys.}

\newcommand{\solphys}{Sol.\ Phys.}
\newcommand\ion[2]{#1$\;${\small\rmfamily{\textsc{\romannumeral #2}}}\relax}

\begin{document} 

\title{Natural-linewidth measurements of the 3C and 3D soft-x-ray transitions in Ni \textsc{xix}}

\author{Chintan~Shah}\email{chintan.shah@mpi-hd.mpg.de}
\affiliation{NASA/Goddard Space Flight Center, 8800 Greenbelt Rd, Greenbelt, MD 20771, USA}%
\affiliation{Max-Planck-Institut f\"ur Kernphysik, Saupfercheckweg 1, 69117 Heidelberg, Germany}%
\affiliation{Department of Physics and Astronomy, Johns Hopkins University, Baltimore, MD 21218, USA}

\author{Steffen~ K\"uhn}
\affiliation{Max-Planck-Institut f\"ur Kernphysik, Saupfercheckweg 1, 69117 Heidelberg, Germany}%
\affiliation{Heidelberg Graduate School of Fundamental Physics, Ruprecht-Karls-Universit\"at Heidelberg, Im Neuenheimer Feld 226, 69120 Heidelberg, Germany}

\author{Sonja~Bernitt}
\affiliation{Max-Planck-Institut f\"ur Kernphysik, Saupfercheckweg 1, 69117 Heidelberg, Germany}%
\affiliation{Institut f\"ur Optik und Quantenelektronik, Friedrich-Schiller-Universit\"at Jena,  Max-Wien-Platz 1, 07743 Jena, Germany}%
\affiliation{Helmholtz-Institut Jena, Fr\"obelstieg 3, 07743 Jena, Germany}%
\affiliation{GSI Helmholtzzentrum f\"ur Schwerionenforschung, Planckstra{\ss}e 1, 64291 Darmstadt, Germany}%

\author{Ren\'e~Steinbr\"ugge}
\affiliation{Deutsches Elektronen-Sychrotron DESY, Notkestra{\ss}e 85, 22607 Hamburg, Germany}%

\author{Moto~Togawa}%
\affiliation{Max-Planck-Institut f\"ur Kernphysik, Saupfercheckweg 1, 69117 Heidelberg, Germany}%
\affiliation{Heidelberg Graduate School of Fundamental Physics, Ruprecht-Karls-Universit\"at Heidelberg, Im Neuenheimer Feld 226, 69120 Heidelberg, Germany}

\author{Lukas~Berger}
\affiliation{Max-Planck-Institut f\"ur Kernphysik, Saupfercheckweg 1, 69117 Heidelberg, Germany}%

\author{Jens~Buck}
\affiliation{Institut f\"ur Experimentelle und Angewandte Physik (IEAP), Christian-Albrechts-Universit\"at zu Kiel, Leibnizstr. 11-19, 24098 Kiel, Germany}

\author{Moritz~Hoesch}
\affiliation{Deutsches Elektronen-Sychrotron DESY, Notkestra{\ss}e 85, 22607 Hamburg, Germany}%

\author{J\"orn~Seltmann}
\affiliation{Deutsches Elektronen-Sychrotron DESY, Notkestra{\ss}e 85, 22607 Hamburg, Germany}%

\author{Mikhail~G.~Kozlov}
\affiliation{St.\ Petersburg Electrotechnical University ``LETI'', Prof.\ Popov Str.\ 5, St.\ Petersburg, 197376, Russia}

\author{Sergey~G.~Porsev}
\affiliation{Department of Physics and Astronomy, University of Delaware, Newark, Delaware 19716, USA}

\author{Ming~Feng~Gu}
\affiliation{Space Science Laboratory, University of California, Berkeley, CA 94720, USA}%

\author{F.~Scott~Porter}
\affiliation{NASA/Goddard Space Flight Center, 8800 Greenbelt Rd, Greenbelt, MD 20771, USA}%

\author{Thomas~Pfeifer}
\affiliation{Max-Planck-Institut f\"ur Kernphysik, Saupfercheckweg 1, 69117 Heidelberg, Germany}%
   
\author{Maurice~A.~Leutenegger}
\affiliation{NASA/Goddard Space Flight Center, 8800 Greenbelt Rd, Greenbelt, MD 20771, USA}%

\author{Charles~Cheung}
\affiliation{Department of Physics and Astronomy, University of Delaware, Newark, Delaware 19716, USA}

\author{Marianna~S.~Safronova}
\affiliation{Department of Physics and Astronomy, University of Delaware, Newark, Delaware 19716, USA}
 
\author{Jos\'e~R.~{Crespo~L\'opez-Urrutia}}%
\affiliation{Max-Planck-Institut f\"ur Kernphysik, Saupfercheckweg 1, 69117 Heidelberg, Germany}%

\date{\today}

\begin{abstract}
We used the monochromatic soft X-ray beamline P04 at the synchrotron-radiation facility PETRA III to resonantly excite the strongest $2p-3d$ transitions in neon-like \ion{Ni}{19} ions, $[2p^6]_{J=0} \rightarrow [(2p^5)_{1/2}\,3d_{3/2}]_{J=1}$ and $[2p^6]_{J=0} \rightarrow [(2p^5)_{3/2}\,3d_{5/2}]_{J=1}$, respectively dubbed 3C and 3D, achieving a resolving power of 15,000 and signal-to-background ratio of 30. We obtain their natural linewidths, with an accuracy of better than 10\%, as well as the oscillator-strength ratio $f(3C)/f(3D)$ = 2.51(11) from analysis of the resonant fluorescence spectra. These results agree with those of previous experiments, earlier predictions, and our own advanced calculations.
\end{abstract}

\maketitle
%
\section{Introduction}
%
Because of their closed-shell $n=2$ ground-state configuration, neon-like ions are prevalent in plasmas across a broad range of temperatures. Their strong spectral lines provide a wealth of diagnostic capabilities, including temperature, density, optical depth, and ultraviolet field intensity. Their spectra were extensively studied in investigations of the Sun and other celestial bodies with the \textit{Chandra} and \textit{XMM-Newton} observatories~\cite{parkinson1973new, smith1985, schmelz1992, waljeski1994, phillips1996, Mauche2001, behar2001chandra, doron2002, xpb2002, gu2003FeL, pfk2003, pradhan2011atomic, beiersdorfer2018, gu2019, gu2020capellaEBIT, grell2021}. The spectrum of Fe XVII is the most commonly studied in the soft x-ray band due to the high cosmic abundance of iron and the easily accessible energy of its $n=2-3$ transitions, while Ni XIX is the second-most abundant such species.

Despite the apparent simplicity of their electronic structure, systematic discrepancies in the intensities of their strongest emission lines have long been noted between theory, astrophysical observations, and laboratory measurements~(see~\cite{brown2008brief} and references therein). Recently, we solved for Ne-like Fe XVII a problem that persisted for several decades: the oscillator-strength ratio between its $2p-3d$ resonance (3C: $[2p^6]_{J=0} \rightarrow [(2p^5)_{1/2}\,3d_{3/2}]_{J=1}$) and intercombination (3D: $[2p^6]_{J=0} \rightarrow [(2p^5)_{3/2}\,3d_{5/2}]_{J=1}$) lines consistently deviated from predictions~\cite{brown2008brief}. Previous experiments on this topic suffered from both known and unexpected systematic uncertainties~\cite{brown1998laboratory, brown2001diagnostic, brown2001systematic, brown2006energy, beiersdorfer2002laboratory, gillaspy2011fe, beiersdorfer2004laboratory, beiersdorfer2017, shah2019, bernitt2012unexpectedly, ock2014, Loch_2015, kuhn2020high}. Our resonant excitation scheme using synchrotron radiation with a much higher resolution and signal-to-noise ratio finally brought experiment and theory into agreement~\cite{kuhn2022}.

In most measurements of Fe XVII, an inner-shell satellite from Na-like ions (line C: $[2p^6 3s]_{J={1/2}}\rightarrow[(2p^5)_{1/2} (3s 3d)_{5/2}]_{J={3/2}}$) blended with the Ne-like line 3D, leading to potential systematic errors in the line ratio~\cite{brown2001systematic}. Experiments in which 3C and 3D were excited by electron impact, and which had sufficient spectral resolving power to detect other lines of Na-like Fe, allowed correction of the strength of line 3D for contamination from line C, as well as optimization of the experimental conditions to minimize production of that undesired charge state. Photoexcitation measurements suffer from a second, subtler effect: the strong autoionizing branch of the upper level of line C continuously feeds the Ne-like ground state. In these experiments with insufficient resolution to split lines C and 3D, the resulting population transfer severely affected the apparent 3C/3D line-intensity ratio~\cite{wu2019change}. Only the most recent photoexcitation measurements mentioned above~\cite{kuhn2022}, with a resolving power of 20,000, could reduce these detrimental transfer effects, reducing the effective overlap between lines C and 3D to less than 1\%. 

Another approach to understanding issues with the 3C/3D ratio in Ne-like iron is to study the same ratio along the isoelectronic sequence as a function of the atomic number $Z$. This has two advantages: first, any systematic errors peculiar to a single experiment are revealed as outliers; second, the scaling with the atomic number of any deviation from predictions can guide future theoretical investigations. The 3C/3D ratios for Ne-like ions ranging from Cr XV to Kr XXVII ($Z=24-36$) were measured at the Lawrence Livermore National Laboratory (LLNL) using an electron beam ion trap (EBIT) equipped with crystal spectrometers~\cite{brown2002}, showing systematic departures from theory of 10--20\%. Extending our photoexcitation experiment to Ne-like Ni, where the lines 3D and C are much farther apart than in Fe XVII, would fully suppress the undesired 3D-C overlap. This, the closeness of nickel to iron in $Z$, and the astrophysical importance of nickel motivated our present measurement in Ne-like Ni XIX ions. Because of the lower chemical abundance of nickel, its L-shell lines are weaker than those of iron in astrophysical sources. Ni XIX is, nevertheless, extremely useful for understanding the spectra of the solar and stellar coronae~\cite{loulergue1975ni,hutcheon1976ni} and for determinations of abundances, as well as plasma temperature and density~\cite{gu2004}. In many high-energy-density plasmas~\cite{rogers1994,seaton1994,beiersdorfer1996observation}, 3C and 3D are stronger than the other L-shell transitions and affect the Rosseland mean opacity, for which recent studies at temperatures akin to stellar interiors disagreed by 10-20\% from models for Ni~\cite{nagayama2019}. Thus, an experimental benchmark is also needed for validating the underlying atomic data in opacity models, and could help to clarify the iron-opacity problem~\cite{bailey2007,fontes2015}.

The ability of atomic methods to accurately predict core parameters, such as transition energies, transition rates, and subsequently derived values, e.g., collisional cross sections, critically depends on the benchmark tests in which predicted values are compared against the experiment. The most readily accessible properties are transition energies, for which the most accurate experimental data can be obtained. Comparing theoretical and experimental energies is an excellent start to testing the atomic methods, as discrepancies in the energies immediately point to the method deficiencies. However, the agreement of the energies does not predict a high level of accuracy of all the other properties. Energy comparisons do not account for different dependences of the atomic properties on the distance from the nucleus, additional correlation corrections specific to the transition operators, and subtler effects of the configuration mixing. Thus, to fully validate atomic structure calculations, experiments measuring natural linewidths and lifetimes are essential.

Lifetimes of excited highly charged ions (HCIs) from the optical to the X-ray domain have been measured for decades using among others things, accelerators, storage rings, and EBITs. At accelerators, beam-foil techniques cover a range of a few nanoseconds to hundreds of femtoseconds~\cite{betz197,Traebert2005PhysScr,traebert2010JPB}, relying on spatially resolving X-ray emission following the passage of a fast ion through a thin foil. However, complexities arising from multiple excitations and the charge-state distribution limits in general the accuracy of such data. Kingdon traps were also used for some lifetime measurements on ions~\cite{moehs2000,moehs2001,smith2005PRA}. Storage rings have allowed for many accurate lifetime studies up to the range of seconds, e.~g.~in Refs.~\cite{mannervik1997PRA,traebert2002JPB,traebert2002CJP,traebert2003JPB,traebert2004JPB,traebert2006JPB,traebert2012PRA,traebert2012NJP,traebert2012JPB, elmar2022,elmar2023} upon injection of excited HCIs from external sources \cite{habs1989first, doerfert1997}, and in the optical range also using re-excitation of the circulating ions with lasers \cite{klaft1994,seelig1998,lochmann2014PRA}. In EBITs, lifetimes ranging from milliseconds down to nanoseconds are accurately measured by monitoring the decay of fluorescence after pulsed excitation, yielding uncertainties as low as 0.15\%~\cite{crespo1998,beiersdorfer2003APJ, crespo2006,crespo2010,beiersdorfer2016}. For the femtosecond range, natural linewidths were accessed with high-resolution crystal spectrometers using standalone EBITs~\cite{beiersdorfer1996observation}, or combined with synchrotron-radiation excitation~\cite{rudolph2013x,steinbrugge2015absolute,steinbruegge2022}. However, modeling line profiles remains difficult, substantially limiting the achievable accuracy. 

In this paper, we present resonant photoexcitation of trapped Ni XIX ions at the monochromator beamline P04 at the PETRA III facility, with a focus on the two strong emission lines 3C and 3D. Several experimental improvements increased our resolving power to $\sim$15,000 and the signal-to-noise ratio to $\sim$30, enabling accurate determination of the 3C/3D oscillator-strength ratio with approximate statistical and total uncertainties of 0.5\% and 4.5\%, respectively. Our observed line shapes yield the absolute natural linewidths, lifetimes, and oscillator strengths for the 3C and 3D lines with an accuracy better than 10\%. All measurements agree well with our predictions.

%
\section{Measurements and Analysis}
%
\begin{figure*}
    \centering
    \includegraphics[width=0.85\textwidth]{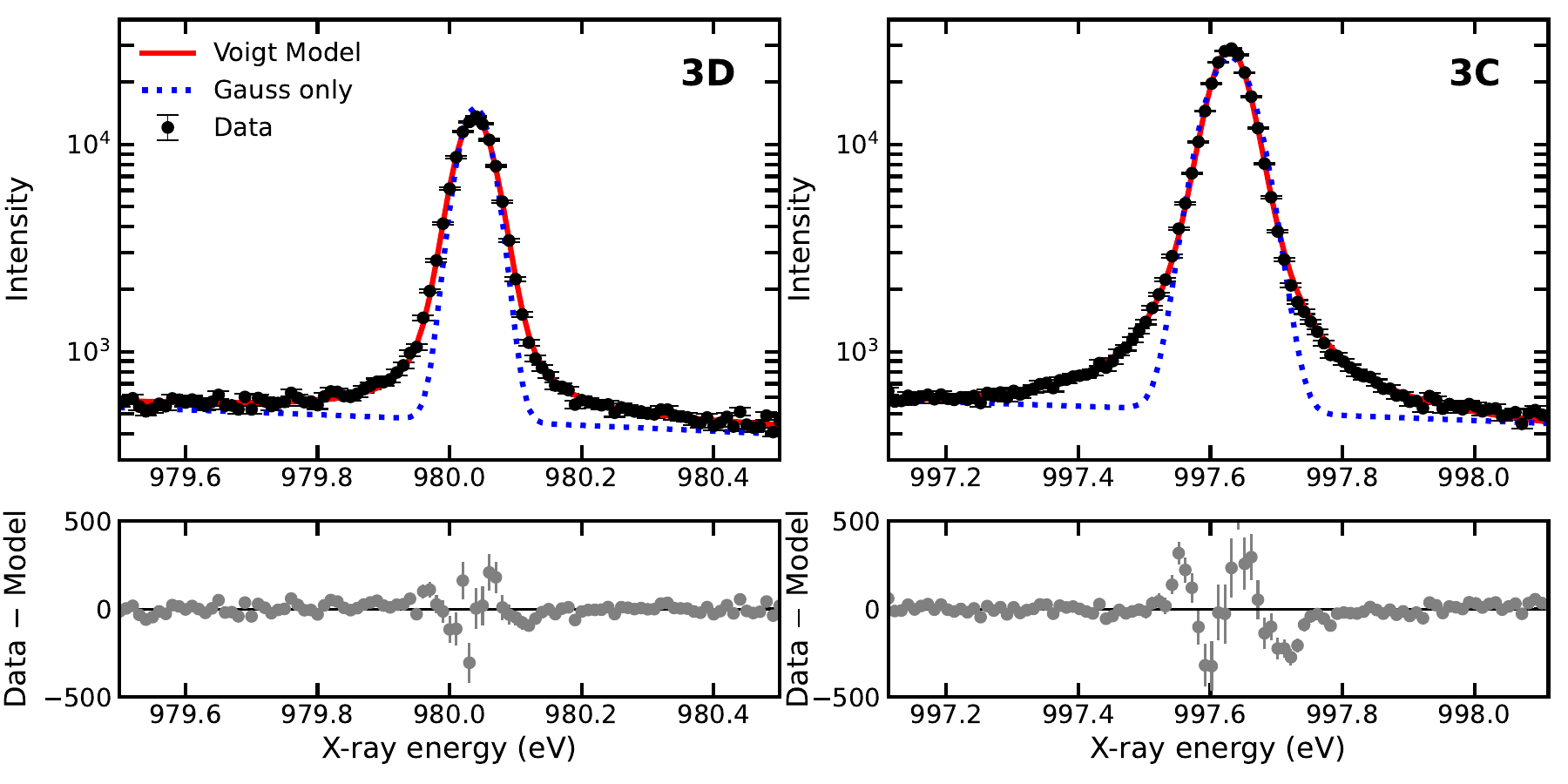}
    \caption{
    Summed fluorescence yield of the 3C and 3D soft X-ray transitions of Ni XIX versus excitation photon energy. Fitted Voigt profiles are shown in red, with their residuals in the bottom panels. 
    {A slight asymmetry of the peak causes the residuals near the center of the peak to deviate from zero. }
    For comparison, blue curves show only the Gaussian component of the model, normalized to the same peak intensity. 
    }
    \label{fig:spectra}
\end{figure*}

We carried out the experiment with PolarX-EBIT~\cite{micke2018}, which is dedicated to the study of interactions of trapped HCIs with photons from external sources, at the P04 beamline~\cite{viefhaus2013variable} of the PETRA III synchrotron-radiation facility. An off-axis electron gun emits a nearly mono-energetic electron beam that is compressed to a diameter of less than 100~$\mu$m by an 870~mT magnetic field generated by an array of permanent magnets. Precursor atoms of Ni were brought in as a tenuous molecular beam of nickel\-ocene (bis(cyclopentadienyl)nickel $(\mathrm{C}_5\,\mathrm{H}_{5})_{2}\,\mathrm{Ni}$) injected into the trap region through a two-stage differential pumping system. Successive electron impacts ionize Ni atoms to the charge state of choice. The ions are confined radially by the negative space-charge potential produced by the $\sim$5.5~mA and $\sim$1.1~keV electron beam, and axially by the $\sim$10~V potential difference given to the drift tubes before and after the central one. At the soft X-ray beamline P04, an APPLE II undulator generates circularly polarized photons which are monochromatized with a variable line-spacing platinum-coated 1200 lines/mm grating~\cite{viefhaus2013variable}. A Kirkpatrick-Baez (KB) mirror system refocuses the photon beam at the position of the PolarX trap region a few meters downstream. The focus diameter there is slighlty smaller than the ion cloud itself, which is approximately 200 micrometers wide. The photon beam enters PolarX from the side where the off-axis electron gun is mounted, and propagates along its longitudinal axis, defined by the magnetic field and the narrow electron beam that it guides and compresses. The photon beam is focused for maximum overlap onto the approximately 16 mm long cloud of Ni ions confined within the central trap electrode. Upon excitation, the resulting fluorescence is detected by two identical silicon drift detectors (SDD) with a resolving power of 10\% at 1~keV mounted at right angles to the photon beam on the top and the side of the trapping region. After passing the trap region, the photon beam exits PolarX unimpeded through its collector, and enters a downstream beamline, where we measure its intensity. 

For production of Ni XIX ions, the electron-beam energy must exceed 607~eV. This also leads to electron-impact-induced fluorescence from dielectronic recombination as well as resonant and direct excitation~\cite{shah2019} that is much stronger than the sought-after photoexcitation signal. We address this by periodically switching within a few microseconds the electron-beam energy between two optimized values, breeding Ni XIX ions at 1090 eV for 200 ms and detecting their fluorescence at 280 eV for 50 ms. This suppresses in-band background photons induced by electron-impact processes, resulting in a cleaner photoexcitation signal~\cite{kuhn2022}. The residual background results from electron impact production of few-hundred-eV photons which partly blend in the SDD with 3C and 3D due to its finite energy resolution and the width of the region of interest. Optimization of the ion-breeding duty cycle allowed us to further narrow the monochromator exit slit width to 25~$\mu$m while keeping a strong fluorescence signal and achieving a robust signal-to-background ratio close to $\sim$30. The narrow slit allows the resolving power of the monochromator to reach values up to 15,000 for the Ni XIX 3C and 3D transitions. 

We scanned a range of $\pm$500 meV around the known centroid energies~\cite{gu2004,gu2007} in 100 steps of 10 meV each, while integrating the photoexcitation signal for $\sim$6 s at each monochromator step to ensure sufficient statistics in the line wings. To generate the spectrum, X-rays detected within a given region of interest centered on the expected energies of 3C and 3D were summed and projected onto the monochromator energy axis. These scans were repeated 20 times, and the resulting (summed) data are shown in Fig.~\ref{fig:spectra}. Each individual scan is fitted with a Voigt profile, a convolution of Gaussian and Lorentzian distributions. The Gaussian component arises from the Doppler width of the trapped ions and the apparatus profile of the monochromator~\citep{hoesch2022}. Meanwhile, the Lorentzian component, as explained in detail in our previous work~\citep{kuhn2022,shah2024}, results not only from the natural linewidth of the transitions but also from a pseudo-Lorentzian instrumental component caused by X-ray diffraction at beamline components~\cite{follath2010}. Total intensities were determined from the area under the curve, derived through a maximum-likelihood fit of Voigt profiles using the Cash statistic~\cite{cash1979,kaastra2017}. Furthermore, 3C and 3D intensities were corrected for the presence of a 500-nm Al filter in front of SDDs, and normalized by the photon flux measured downstream of the EBIT with a calibrated photodiode, which together increase the ratio by 0.5\%. Given that the intensity of the 3C and 3D transitions excited by the monochromatic X-ray beam is directly proportional to the oscillator strength of each transition~\cite{bernitt2012unexpectedly,traebert2022}, we derived an oscillator strength ratio of 2.51(2) from the measured intensities. We note that the inner-shell satellite C of the Na-like ion, previously a major source of systematic uncertainties in many experiments on Fe XVII, does not affect the Ni XIX 3C/3D line ratio. This is primarily because line C is clearly separated from line 3D, falling well outside the scan range we used. From our low-resolution measurement with a 1-mm exit slit, we determined the difference $\Delta{{E}}_{\mathrm{3D}-\mathrm{C}}\approx2.4$~eV. 

\begin{figure}
    \centering
    \includegraphics[width=\columnwidth]{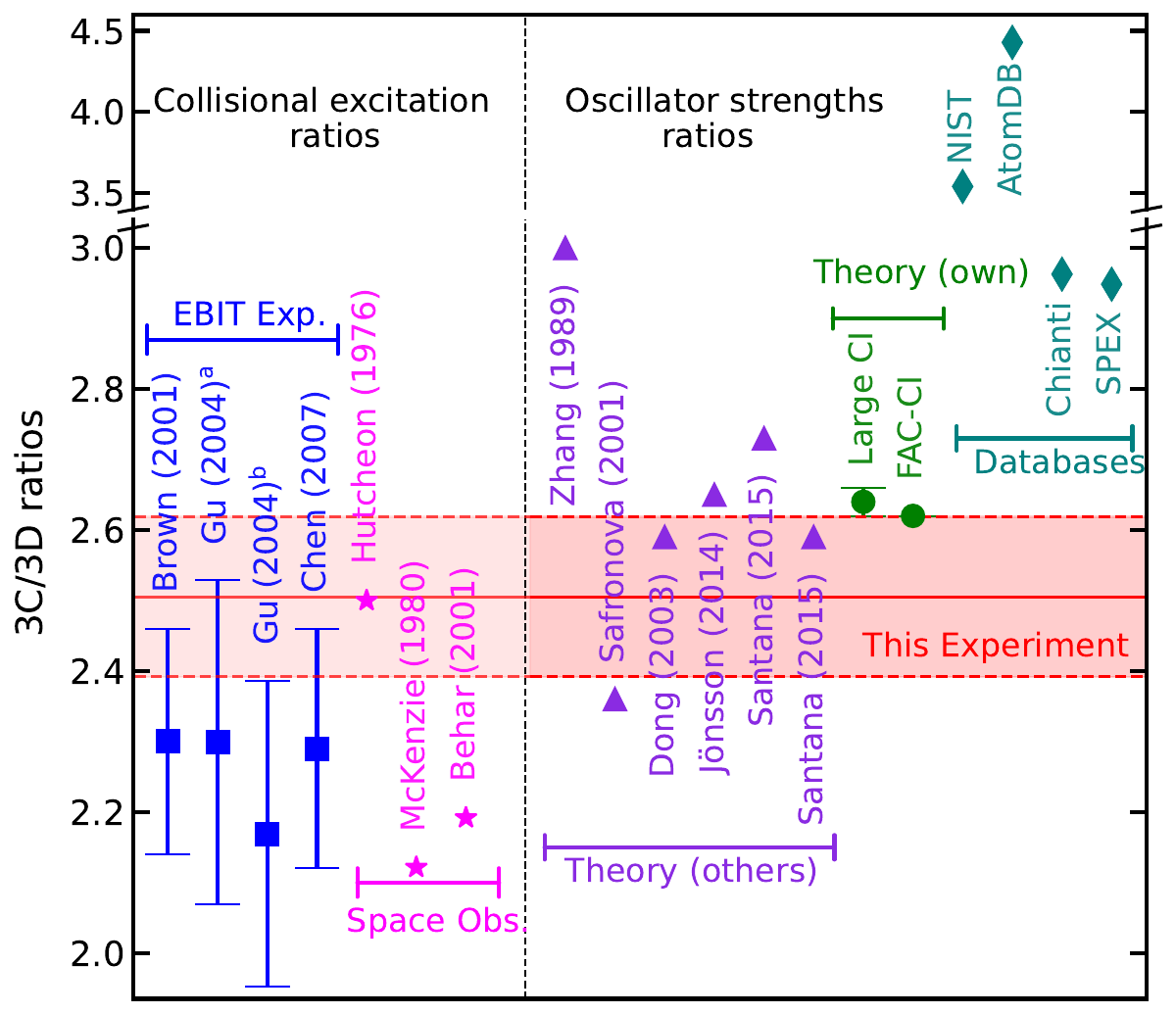}
    \caption{Present experimental Ni XIX 3C/3D oscillator-strength ratio in comparison with collision-strength ratios measured in prior EBIT experiments: experiments with a crystal spectrometer in~\citet{brown2001systematic}, with a grating spectrometer in \citet{gu2004laboratory} (denoted with a superscript ${a}$), and with microcalorimeters in \citet{gu2004laboratory} (denoted with a superscript ${b}$) and \citet{chen2007prl}; solar observations~\cite{hutcheon1976ni,mckenzie1980solar}; and Chandra studies of Capella~\cite{behar2001chandra}. Our measured oscillator-strength ratio is also compared with earlier theoretical studies~\cite{zhang1989relativistic,safronova2001,dong2003,jonsson2014,santana2015electron}, our own predictions~\cite{cheung2021scalable,gu2008flexible}, and commonly used spectral plasma models and databases~\cite{NIST_ASD,foster2012,delzanna2009,kaastra1996spex}.
    }
    \label{fig:results_ratio}
\end{figure}

The measured line ratio can be affected by periodic fluctuations of the actual photon energy around its nominal value, stemming from incorrect interpolation tables for the grating and mirror angular encoders~\citep{follath2010,Krempasky2011} of the monochromator, as discussed in our previous work~\cite{kuhn2022,shah2024,togawa2023}. To avoid this issue in subsequent works, we simultaneously measured a proxy for the fluctuations of the true photon energy with a photoelectron-energy spectrometer (ASPHERE: Angular Spectrometer for Photoelectrons with High Energy REsolution)~\citep{ROSSNAGEL2001}. Unfortunately, this instrument was not available during the present measurements, but in follow-up campaigns~\cite{shah2024} we found with it oscillating differences between nominal and true photon energies of up to $\approx\pm70~\mathrm{meV}$ in the Ni XIX 3C and 3D scan range. We simulated such monochromator fluctuations with mock 3C and 3D Voigt profiles, resulting in systematic uncertainties of approximately 3\% in the 3C/3D intensity ratio. A slight line-profile asymmetry caused by X-ray diffraction at beamline components was quantified by fitting with skewed Voigt profiles, showing changes in the ratio of less than 0.5\%. We observed scan-to-scan variations in the amplitudes of 3C and 3D of $\sim$3\% which are not attributable to any known cause, and which we therefore take as a term in our systematic error budget. All above systematic uncertainties, including the $\sim$0.1\% uncertainty in the 3C/3D ratio arising from a 10\% uncertainty in the thickness of the 500-nm Al optical filter, have been taken into account in the final error budget for the inferred oscillator-strength ratio, namely $f(3C/3D) = 2.51(2)_\mathrm{stat}(11)_\mathrm{sys}$, as displayed in Fig.~\ref{fig:results_ratio}. Note that the circular polarization of the photon beam does not influence the ratio, as both transitions 3C and 3D have identical angular emission characteristics (see Appendix A in Ref.~\cite{steinbruegge2022})

To determine the natural linewidths of 3C and 3D, we used the Gaussian and Lorentzian widths extracted through Voigt fits to twenty scans of 3C and 3D. While Gaussian widths of approximately 50 meV for 3C and 3D were consistent with the experimental conditions, the extracted Lorentzian widths of $\sim$30 and $\sim$17 meV for 3C and 3D transitions, respectively, were clearly larger than expected from theory. We attribute this discrepancy to an additional pseudo-Lorentzian component induced by X-ray diffraction at beamline components \cite{follath2010,Krempasky2011} that artificially raises the apparent Lorentzian linewidth of observed transitions, as shown in our previous work~\cite{kuhn2022}. We checked that here by measuring the K$\alpha$, K$\beta$, and K$\gamma$ X-ray transitions from helium-like F VII and Ne IX several times. Their theoretical natural linewidths were taken from the NIST Atomic Spectral Database (ASD)~\cite{NIST_ASD,cann1992,johnson2002}, and we assigned them a conservative 10\% uncertainty~\cite{gu2004}. As Lorentzian contributions add linearly, in contrast to the quadratic addition of Gaussian widths, we subtracted the theoretical natural linewidths from the Lorentzian linewidths inferred from the measurements. We noticed an increase in the pseudo-Lorentzian beamline component dependent on the monochromator energy, which we model empirically as a quadratic function, determining the beamline Lorentzian contribution at 3D and 3C line energies to be 10.0(1.0) and 10.3(1.1) meV, respectively. We also derived beamline Lorentzian contributions of 7.2(5) and 7.4(5) meV at Fe XVII 3D and 3C energies, respectively, and upon comparison with the 7.0(3) meV obtained in our earlier work~\cite{kuhn2022}, which relied exclusively on the single F VIII K$\beta$ line, we find reasonably good agreement between the present work and~\citet{kuhn2022}. 

\begin{figure*}
    \centering
    \includegraphics[width=\textwidth]{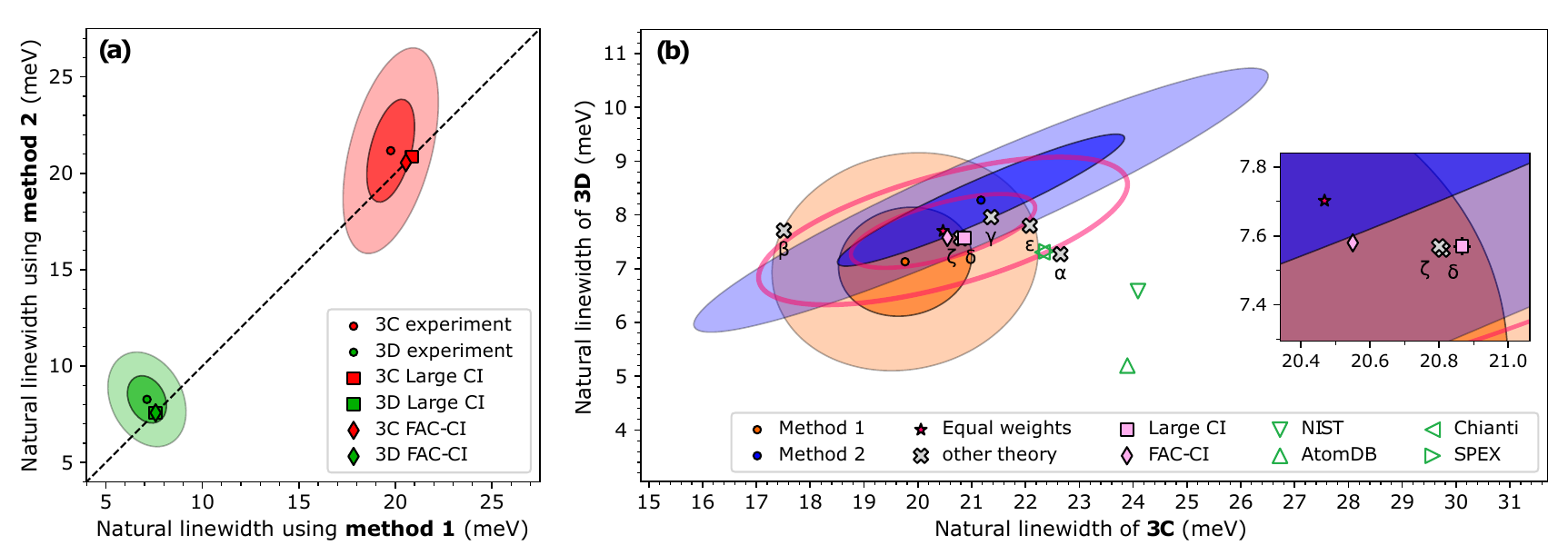}
    \caption{(a) Results for the natural linewidths of 3C (red) and 3D (green) for Ni~XIX analyzed using two different methods, with ellipses displaying their one- and two-sigma experimental uncertainties. The dashed diagonal line marks where the two methods coincide. Predictions from large-scale configuration interaction (Large-CI; squares) and Flexible Atomic Code Configuration Interaction (FAC-CI; diamonds) are shown. (b): Natural line widths inferred for 3C and 3D. Ellipses: one- and two-sigma experimental uncertainties for method 1 (orange) and method 2 (blue), as well as for their mean with equal weights (magenta star and lines). Symbols: Predictions from our large-CI (pink square), our FAC-CI (pink diamond), other calculations (gray crosses: $\alpha$ Distorted Wave (DW),~\citet{zhang1989relativistic}, $\beta$ many-body perturbation theory (MBPT),~\citet{safronova2001}; $\gamma$ multi-configuration Dirac Fock (MCDF):~\citet{dong2003}, $\delta$ (CI),~\citet{jonsson2014}, and~$\varepsilon$ (CI), $\zeta$ (MBPT)~\citet{santana2015electron}.}
    \label{fig:results_gamma}
\end{figure*}

To derive the natural linewidths of Ni XIX 3C and 3D, we subtracted the pseudo-Lorentzian instrumental component from the Lorentzian widths obtained from Voigt fits to each scan of both lines. The determined natural linewidths were 19.8(1.2) and 7.1(1.0) meV for 3C and 3D, respectively, indicated as "method 1" in Tab.~\ref{tab:gammas} and in Fig.~\ref{fig:results_gamma}. Their uncertainties include the statistical error on individual widths obtained from the fit, and systematic uncertainties from asymmetric line shapes (2\%) and monochromator energy fluctuations at 3C (2.2\%) and 3D (2.3\%) and at the helium-like K-shell reference transitions of helium-like F and Ne ions (6\%). 

\begin{table}[b]
  \centering
  \caption{Natural linewidths ($\Gamma^\mathrm{exp}$) for 3C and 3D of Ni~XIX in meV, as determined using two different methods, as well as their mean. Pearson correlation coefficients $\rho_\mathrm{3C,3D}$ for $\Gamma_\mathrm{3C}$ and $\Gamma_\mathrm{3D}$ are also listed.}
    \begin{tabular}{@{\extracolsep{\fill}} lccc}
    \hline \hline
    Natural linewidths ($\Gamma^\mathrm{exp}$) & Line 3C (meV)& Line 3D (meV)& $\rho_\mathrm{3C,3D}$ \\
    \hline
    Method 1 & 19.8\,$\pm$\,1.2 & 7.1\,$\pm$\,1.0 & 0.10 \\
    Method 2 & 21.2\,$\pm$\,2.7 & 8.3\,$\pm$\,1.2 & 0.94 \\
    \midrule
    Mean (equal weights) & 20.5\,$\pm$\,1.7 & 7.7\,$\pm$\,0.7 & 0.63 \\
    \hline \hline
    \end{tabular}%
  \label{tab:gammas}%
\end{table}%

Alternatively, we can use the following equations to determine the individual natural linewidths of 3C and 3D using the {\it difference} in their uncalibrated Lorentzian widths, as shown in~\citet{kuhn2022}:
\begin{equation*}
	\Gamma_{\mathrm{3C}} =
	\frac{\Delta\Gamma_{\mathrm{{3C-3D}}}}{1-{f ( \frac{\mathrm{{3D}}}{\mathrm{{3C}}}) E ( \frac{\mathrm{{3D}}}{\mathrm{{3C}}})^2 }},
	\,\, \mathrm{and\,\,\,}
	\Gamma_{\mathrm{3D}} =
	\frac{\Delta\Gamma_{\mathrm{{3C-3D}}}}{{f ( \frac{\mathrm{{3C}}}{\mathrm{{3D}}}) E ( \frac{\mathrm{{3C}}}{\mathrm{{3D}}})^2 }-1}.
	\label{3c_3d_gamma}
\end{equation*}
However, in this second method, we assume that the pseudo-Lorentzian contribution from the beamline is similar for both 3C and 3D lines, as in~\citet{kuhn2020high}. This allows the subtraction of Lorentzian widths obtained directly from the 3C and 3D Voigt fits ($\Delta \Gamma_{\mathrm{3C} - \mathrm{3D}} = 13.0(1.4)$ meV). By utilizing this difference along with the measured $f$(3C/3D) oscillator-strength ratio in the present work and the transition-energy $E$(3C/3D) ratio measured in \cite{gu2007} in the above equations, we determine the natural linewidths of 3C and 3D to be 21.2(2.7) and 8.3(1.2) meV, respectively. The final errors in this method (method 2) account for Voigt fit statistical errors on individual widths, asymmetric line shapes (2\%), and the effects of monochromator energy fluctuations on 3C (2.2\%) and 3D (2.3\%), and on $\Delta \Gamma_{\mathrm{3C} - \mathrm{3D}}$ (5\%), in addition to the uncertainty in $f({\mathrm{3C}/\mathrm{3D}})$ from this work and the $E({\mathrm{3C}/\mathrm{3D}})$ uncertainty from \cite{gu2007}.

In both methods, we propagated uncertainties and covariances using Monte-Carlo methods, obtaining final natural linewidths values for the 3C and 3D lines as unweighted means of all 20 individual measurements. A comparison of the results of both methods is displayed in Fig.~\ref{fig:results_gamma}(a). Figure~\ref{fig:results_gamma}(a) shows that results for both linewidths are largely uncorrelated in method 1, as it is less prone to systematics, while method 2 exhibits a Pearson correlation coefficient close to 1 for both linewidths, indicating high correlation and greater susceptibility to systematics arising from amplitude ratios, energy ratios, and the assumption of similar beamline components. Since both methods use essentially the same data set but produce slightly different results, we use equal weighting to ensure a balanced representation of both methods for our final results presented in Tab.~\ref{tab:gammas} and shown as magenta ellipses in Fig.~\ref{fig:results_gamma}(b). The mean uncertainties given in Tab.~\ref{tab:gammas} are the 1-sigma widths of the projections of the ellipses onto their respective natural linewidth axes as represented by the magenta ellipses in Fig.~\ref{fig:results_gamma}(b).

%
\section{Results and Discussions}

Figure~\ref{fig:results_ratio} compares our results with earlier experimental data, observations, and predictions. Our own calculations using large-scale configuration interaction (large-CI)~\cite{cheung2021scalable} and FAC-CI~\cite{gu2008flexible} methods, both accounting for the relativistic Breit interaction and quantum electrodynamics (QED) effects, yield an oscillator-strength ratio which agrees with our experiment within the uncertainties. Details of large-CI computations are given in the Appendix~\ref{app}. Additional comparisons with other theoretical values from the literature showed typical departures within a 1--2$\sigma$ range, with the largest one being from \citet{zhang1989relativistic}. 

Three laboratory measurements of the 3C/3D intensity ratio of Ni XIX based on electron-impact excitation were reported~\cite{brown2002,gu2004,chen2007prl}. The collisional excitation ratios from all three previous experiments span 1.90--2.35, slightly lower than the oscillator strength ratio measured in our experiment, but all four experiments are mutually consistent within uncertainties. Observations of the solar corona \cite{hutcheon1976ni, mckenzie1980solar} and Capella \cite{behar2001chandra} show some scatter in the 3C/3D ratio, but in the absence of uncertainty estimates we cannot quantify the level of agreement with laboratory measurements.

Figure~\ref{fig:results_gamma} (a) shows a comparison between measured 3C and 3D natural linewidths obtained using two different methods, which agree (see dashed diagonal lines) with each other and our predictions. Our calculations using both large-CI and FAC-CI methods agree within the 1$\sigma$ uncertainty of our experimental results. As can be seen in Figure~\ref{fig:results_gamma} (b), the other older calculations from~\citet{zhang1989relativistic} and~\citet{safronova2001} show disagreement; however, their values fall just outside of the 2-sigma ellipse of mean natural linewidths, whereas somewhat recent calculations presented in Refs.~\cite{dong2003,jonsson2014,santana2015electron} show agreement within the 1-sigma ellipse.

The individual oscillator strengths of 3C and 3D can be derived from the natural linewidth using the relation $f_{fi}^\mathrm{exp} = C \,\, \lambda^2_{if} \,\, (g_i/g_f) \,\, (
\Gamma_\mathrm{exp}/ \hbar)$, where $ C = 1 / (32 \,\pi^3 \,\alpha \,a_0^2 \,Ry) = 1.499\,19 \times 10^{-14}~\mathrm{{nm}^{-2} s}$. $g_i$ and $g_f$ represent the statistical weights of the initial ($i$) and final ($f$) states, respectively. $\lambda_{if}$ is the transition wavelength given in nanometers taken from \cite{gu2007}, and $\hbar$ is the reduced Planck constant. The experimental oscillator strengths for 3C and 3D are found to be 2.17\,$\pm$\,0.18 and 0.84\,$\pm$\,0.08, respectively. This corresponds to lifetimes of 32.11\,$\pm$\,2.68 fs and 85.48\,$\pm$\,7.84 fs for 3C and 3D, respectively. All of these measured quantities agree very well with our predictions. 

Our present experimental validation of oscillator strengths effectively eliminates uncertainties in atomic data as potential contributors to the observed discrepancies in nickel opacity measurements~\cite{nagayama2019} and iron opacity~\cite{bailey2007,fontes2015,kuhn2022}. The measured oscillator strengths offer direct application in astrophysical modeling, enabling the diagnosis of turbulent velocity in moderately optically thick plasmas~\cite{xu2002high,werner2009constraints}. Following a comparing with oscillator strength ratios and natural linewidths from established databases such as NIST-ASD~\cite{NIST_ASD}, AtomDB~\cite{foster2012}, Chianti~\cite{delzanna2009}, and SPEX~\cite{kaastra1996spex}, significant discrepancies were identified with our experimental results. There is a pressing need to update these databases with such precise measurements in view of the crucial role of accurate data in modeling observational spectra from existing missions like \textit{Chandra} and \textit{XMM-Newton}, as well as upcoming observations with \textit{XRISM}~\cite{xrism2018}, which was recently successfully launched, as well as future planned missions such as \textit{Athena}~\citep{pajot2018athena}, \textit{LEM}~\citep{lem2022}, \textit{HUBS}~\citep{hubs2020}, \textit{Arcus}~\citep{arcus_instrum}, \textit{HiReX}~\citep{hirex2021}, and \textit{Lynx}~\citep{Schwartz2019}. The present combination of experimental benchmark and converged calculations is a crucial consistency check for atomic data required to interpret X-ray observations in astrophysics, fusion, and high-energy density plasma research.

%
%
\begin{acknowledgments}
This research was funded by the Max Planck Society (MPG) and the German Federal Ministry of Education and Research (BMBF) under Project No. 05K13SJ2. C.S. acknowledges support from MPG and NASA. F.S.P. and M.A.L. acknowledge support from the NASA Astrophysics Program. The theoretical work was supported by the US NSF Grant No. PHY-2012068 and No. PHY-2309254,  U.S. Office of Naval Research Grant No. N00014-20-1-2513, and the European Research Council (ERC) under the Horizon 2020 Research and Innovation Programme of the European Union (Grant Agreement No. 856415). Calculations were performed in part using the computing resources at the University of Delaware, in particular the Caviness and DARWIN high-performance computing clusters. This theoretical work, including M.G.K.'s contributions, was carried out as part of a scientific collaboration during the calendar year 2020. We acknowledge DESY (Hamburg, Germany), a member of the Helmholtz Association HGF, for the provision of experimental facilities. Parts of this research were carried out at PETRA~III. We thank Jens Viefhaus and Rolf Follath for valuable discussions on X-ray monochromator resolution and performance, and the synchrotron-operation team and P04 team at PETRA~III for their skillful and reliable work. MSS thanks MPIK, Heidelberg, for hospitality. 
\end{acknowledgments}
%
\appendix*
\section{Theory}\label{app}
%
The calculations are carried out using a large-scale configuration interaction (CI) method~\cite{cheung2021scalable}, including correlations from all 10 electrons. Basis sets of increasing sizes are used to check for convergence of the values. The basis set is designated by the highest principal quantum number for each partial wave included. For example, [$12spdfg$] means that all orbitals up to $n = 12$ are included for the $spdfg$ partial waves. We begin by considering all possible single and double excitations to orbitals up to $5spdf6g$ from the $2s^2 2p^6$ and $2s^2 2p^5 3p$ even configurations and $2s^2 2p^5 3s$, $2s^2 2p^5 3d$, and $2s 2p^6 3p$ odd configurations, correlating 8 electrons. We verified that inclusion of the $2s 2p^6 3s$ even and $2s^2 2p^5 4s$, $2s^2 2p^5 4d$ odd configurations as basic configurations has a negligible effect on the energies and relevant matrix elements. The calculated contributions to the energies of Ni XIX are listed in Tab.~\ref{ni_table1}. The results are compared with a revised analysis of the experimental data \cite{AK}. We use $jj$-coupling and NIST-style LS-coupling term designations for comparisons. Contributions to the $E1$ reduced matrix elements $D(3D)$ and $D(3C)$ and the $3C/3D$ oscillator strengths ratios are listed in Table~\ref{ni_table2}. The $E(3C/3D)$ energy ratio is 1.018 and the $f(3C/3D)$ oscillator strength ratio is 2.64(2). 

To assess the impact of triple excitations, we consider a broad range of configurations up to $5spdf6g$. As demonstrated in Tables~\ref{ni_table1} and \ref{ni_table2}, these excitations result in negligible corrections to both energies and matrix elements. Subsequently, we expand the basis set to $[12spdfg]$, leading to a significant improvement in the agreement of energies with experimental values and a minimal shift in the ratio ($-0.008$). Further accounting of contributions from the $1s^2$ shell with the $[12spdfg]$ basis improves agreement with experimental energies, albeit with a marginal contribution to the $3C/3D$ ratio ($-0.006$). A comparison of results for $D(3C)$ and $D(3D)$ obtained in length and velocity gauges reveals only a marginal difference of 0.001 for the $[12spdfg]$ basis. Expanding the basis set to $[17spdfg]$ produces a modest improvement in energies compared to experiment, accompanied by a slight shift in the ratio ($-0.001$). Further expansion to $[22spdfg]$ results in minor corrections to energies, with an even smaller contribution to the ratio ($-0.0005$). The quantum electrodynamic (QED) contributions are included following Ref.~\cite{QED}. The inclusion of QED has a small effect on the individual line energies, however, a considerable contribution to the energy difference of 3C and 3D, see Tab.~\ref{ni_table1}. Furthermore, the ratio is changed by $-0.01$ by accounting for QED.

Additionally, we compute the transition rates for all other transitions contributing to the radiative decay of the 3$C$ and 3$D$ levels. The sums of these rates are small and are listed in Tab.~\ref{ni_table2}. Linewidth values correspond to total transition rates, with uncertainties computed from the largest uncertainties in the 3C and 3D matrix elements, including additional configurations in CI space and QED. Additional uncertainties due to $h$ orbitals for transition rates and linewidths, based on recent studies of Fe$^{16+}$~\cite{shah2024}, are included, and the final uncertainties are computed based on the relative difference.

\begin{table*}
\caption{\label{ni_table1} Contributions to the energies of Ni$^{18+}$ calculated with increased size basis sets and greater numbers of configurations. The results are compared with a revised analysis of the experimental data \cite{AK}. All energies are given in cm$^{-1}$ with the exception of the last row, which shows the difference of the 3C and 3D energies in eV. The basis set is designated by the highest quantum number for each partial wave included. For example, $[12spdfg]$ means that all orbitals up to $n=12$ are included for $spdfg$ partial waves. Contributions from the larger basis sets $[17spdfg]$ and $[22spdfg]$, triple excitations, excitations from the $1s^2$ shells, and QED contributions are given separately. The Diff. column represents the absolute difference between Expt. \cite{AK} and our "Final" predictions. Diff (\%) shows this difference in relative percent. }
\begin{ruledtabular}
\begin{tabular}{lccccccccccccc}
\multicolumn{2}{c}{Configuration}&
\multicolumn{1}{c}{$J$}&
\multicolumn{1}{c}{Expt.~\cite{AK}}&  
\multicolumn{1}{c}{$[5spdf6g]$}&
\multicolumn{1}{c}{Triples}&
\multicolumn{1}{c}{$+[12spdfg]$}&
\multicolumn{1}{c}{$1s^2$}& 
\multicolumn{1}{c}{$+[17spdfg]$}&
\multicolumn{1}{c}{$+[22spdfg]$}& 
\multicolumn{1}{c}{QED}&     
\multicolumn{1}{c}{Final}& 
\multicolumn{1}{c}{Diff.~\cite{AK}}& 
\multicolumn{1}{c}{Diff.~\cite{AK}} \\
\hline
$2p^6   $  &$^1S$          & 0 &  0       &        0 &   0   & 0 &   0&    0 & 0   & 0   & 0       & 0    &        \\
$2p^5 3p$  &$\left(\nicefrac{3}{2},\nicefrac{1}{2}\right)$    & 1 &  7381990 &  7374679 & -206 & 3579& 351& 797	& 372 & 44   & 7379615 &  2375 & 0.03\% \\
$2p^5 3p$  &$\left(\nicefrac{3}{2},\nicefrac{1}{2}\right)$    & 2 &  7409915 &  7403138 & -2   & 2880& 264& 724	& 331 & 29   & 7407364 &  2551 & 0.03\% \\
$2p^5 3p$  &$\left(\nicefrac{3}{2},\nicefrac{3}{2}\right)$    & 3 &  7431735 &  7424692 & -4   & 3023& 261& 738	& 340 & 102  & 7429152 &  2583 & 0.03\% \\
$2p^5 3p$  &$\left(\nicefrac{3}{2},\nicefrac{3}{2}\right)$    & 1 &  7440050 &  7433248 & -11  & 2836& 277& 730	& 332 & 85   & 7437497 &  2553 & 0.03\% \\ [0.5pc]
$2p^5 3s $ &$\left(\nicefrac{3}{2},\nicefrac{1}{2}\right)^o$  & 2 &  7105260 &  7096413 & 17   & 3417& 478& 751	& 350 & 1052 & 7102477 &  2783 & 0.04\%  \\
$2p^5 3s $ &$\left(\nicefrac{3}{2},\nicefrac{1}{2}\right)^o$  & 1 &  7122600 &  7114019 & 15   & 3303& 415& 722	& 337 & 1052 & 7119862 &  2738 & 0.04\%  \\
$2p^5 3s $ &$\left(\nicefrac{1}{2},\nicefrac{1}{2}\right)^o$  & 1 &  7247700 &  7249597 & 14   & 3386& 498& 735	& 345 & 1403 & 7255978 & -8278  & 0.11\%  \\
$2p^5 3d $ &$\left(\nicefrac{3}{2},\nicefrac{3}{2}\right)^o$  & 1 &  7807700 &  7801245 & 19   & 2336& 409& 684	& 363 & 72   & 7805128 &  2572 & 0.03\%  \\
$2p^5 3d $ &$\left(\nicefrac{3}{2},\nicefrac{5}{2}\right)^o$  & 2 &  7825770 &  7819486 & 19   & 1839& 415& 651	& 359 & 87   & 7822856 &  2914 & 0.04\%  \\
$2p^5 3d $ &$\left(\nicefrac{3}{2},\nicefrac{5}{2}\right)^o$  & 4 &  7825280 &  7819623 & 19   & 2044& 410& 677	& 362 & 88   & 7823223 &  2057 & 0.03\%  \\
$2p^5 3d $ &$\left(\nicefrac{3}{2},\nicefrac{3}{2}\right)^o$  & 3 &  7830930 &  7825368 & 17   & 1687& 414& 633	& 353 & 81   & 7828554 &  2376 & 0.03\%  \\
$2p^5 3d $ &$\left(\nicefrac{3}{2},\nicefrac{3}{2}\right)^o$  & 2 &  7847100 &  7841657 & 18   & 1638& 410& 654	& 356 & 86   & 7844819 &  2281 & 0.03\%  \\
$2p^5 3d $ &$\left(\nicefrac{3}{2},\nicefrac{5}{2}\right)^o$  & 3 &  7857640 &  7852407 & 17   & 1705& 407& 630	& 350 & 90   & 7855606 &  2034 & 0.04\%  \\
$2p^5 3d $ &$\left(\nicefrac{3}{2},\nicefrac{5}{2}\right)^o$  & 1 &  7901400 &  7899252 & 3    & 1681& 384& 638	& 352 & 136  & 7902446 & -1046 & 0.01\%  \\
$2p^5 3d $ &$\left(\nicefrac{1}{2},\nicefrac{3}{2}\right)^o$  & 2 &  7972475 &  7967013 & 17   & 1683& 484& 657	& 362 & 432  & 7970647 &  1828 & 0.02\%  \\
$2p^5 3d $ &$\left(\nicefrac{1}{2},\nicefrac{5}{2}\right)^o$  & 2 &  7980810 &  7975017 & 17   & 1933& 475& 667	& 362 & 427  & 7978897 &  1913 & 0.02\%  \\
$2p^5 3d $ &$\left(\nicefrac{1}{2},\nicefrac{5}{2}\right)^o$  & 3 &  7986640 &  7981013 & 16   & 1759& 481& 638	& 356 & 443  & 7984706 &  1934 & 0.02\%  \\
$2p^5 3d $ &$\left(\nicefrac{1}{2},\nicefrac{3}{2}\right)^o$  & 1 &  8041800 &  8040754 & -29  & 1668& 383& 633	& 353 & 372  & 8044132 & -2332 & 0.03\%  \\
$3C-3D$ & (eV)  &    & 17.4074  & 17.5440&-0.0039&-0.0016&-0.0002&-0.0006&-0.0001& 0.0291&	17.5669& &  \\
  \end{tabular}
\end{ruledtabular}
\end{table*}

\begin{table*}
\caption{\label{ni_table2} Contributions to the E1 reduced matrix elements $D(3D)=D(2p^6$~$^1S_0 - 2p^5 3d$~$(3/2,5/2))$ and $D(3C)=D(2p^6$~$^1S_0 -2p^5 3d$~$(1/2,3/2))$ (in a.u.) and the ratio of the respective oscillator strengths $R$ in Ni$^{18+}$. See thecaption of Table~\ref{ni_table1} for designations. $L$ and $V$ rows compared results obtained in length and velocity gauges for the $[12spdfg]$ basis. All other results are calculated using the length gauge. The transition rates and linewidth are listed at the bottom. The total of the other transition rates contributing to the lifetime of the 3C and 3D levels is labeled ``Other transitions''.}
\begin{ruledtabular}
\begin{tabular}{lccrcrcc}
\multicolumn{2}{c}{Basis set}&
 \multicolumn{1}{c}{$D(3C)$}& \multicolumn{1}{c}{$\Delta D(3C)$} &
 \multicolumn{1}{c}{$D(3D)$}& \multicolumn{1}{c}{$\Delta D(3D)$} &
 \multicolumn{1}{c}{$R(3C/3D)$} & \multicolumn{1}{c}{$\Delta R$}  \\
\hline
$[5spdf6g]$   &            &0.30012  &          &  0.18530   &         &  2.670  	&		  \\
              &  $+$Triples&0.29999  & -0.00013 &  0.18530   & 0.00000 &  2.668  	& -0.002  \\
$[12spdfg]$   &     $L$    &0.30031  & 0.00019  &  0.18568   & 0.00038 &  2.662  	& -0.008  \\
              &     $V$    &0.30060  &          &  0.18582   &         &  2.663  	&		  \\
$[12spdfg]$   &  +$1s^2$   &0.30018  & -0.00013 &  0.18581   & 0.00013 &  2.656  	& -0.006  \\
$[17spdfg]$   &            & 0.30032 & 0.00001  &  0.18571   & 0.00003 & 2.662   	& -0.001  \\
$[22spdfg]$   &            & 0.30031 & -0.00001 &  0.18572   & 0.00001 & 2.661   	& -0.0005 \\
QED           &            &         & -0.00012 &            & 0.00028 &         	& -0.010   \\
Final         &            & 0.29993 &          &  0.18613   &         & 2.64(2)      &	      \\ \hline
Recommended transition rate (s$^{-1}$)  &    &  3.168(4)$\times10^{13}$ && 1.148(4)$\times10^{13}$&                \\
Other transitions (s$^{-1}$)  &    &  1.75$\times10^{10}$ && 1.59$\times10^{10}$&                \\
Total rate (s$^{-1}$)  &    &  3.170(4)$\times10^{13}$ && 1.150(4)$\times10^{13}$&                \\
Linewidth (meV)  &    &  20.867(26)
 && 7.570(27) &                \\
  \end{tabular}
\end{ruledtabular}
\end{table*}

%
%
\bibliographystyle{apsrev4-2}
\bibliography{references}

\end{document}